\documentclass[11pt]{article}

\usepackage[utf8]{inputenc}
\usepackage[T1]{fontenc}
\usepackage{lmodern}
\usepackage[margin=1in]{geometry}
\usepackage{amsmath,amssymb}
\usepackage{graphicx}
\usepackage{booktabs}
\usepackage{xcolor}
\usepackage{microtype}
\usepackage{caption}
\usepackage{subcaption}
\usepackage[hidelinks]{hyperref}
\usepackage{orcidlink}

\newcommand{\K}{\mathbf{K}}
\newcommand{\I}{\mathbf{I}}
\newcommand{\W}{\mathbf{W}}
\newcommand{\code}[1]{\texttt{#1}}

\title{\textbf{Evaluating the Robustness of PRNU Source-Camera\\
Attribution to the Fingerprint-Copy Attack:\\
An Open Toolkit with Triangle-Test Detection}}

\author{%
Zeno J.~M.~H.~Geradts\,\orcidlink{0000-0001-5912-5295}\\
\small Netherlands Forensic Institute, The Hague, The Netherlands\\
\small University of Amsterdam, Amsterdam, The Netherlands\\
\small \texttt{z.j.m.h.geradts@uva.nl}%
}
\date{\today}

\begin{document}
\maketitle

\begin{abstract}
\noindent
Photo-Response Non-Uniformity (PRNU) is the de~facto standard trace for
sensor-level source-camera attribution, and a high Peak-to-Correlation-Energy
(PCE) between a questioned image and a camera reference fingerprint is routinely
treated as strong evidence of common origin. The fingerprint-copy attack of
Goljan, Fridrich and Chen (2011) shows that this signal can be forged: an
adversary who estimates a target camera's fingerprint from that camera's
publicly available images can inject a scaled copy of it into an unrelated
photograph, producing a forgery that clears a PCE decision threshold despite
never having been taken by the target device. We present \code{prnucf}, an open,
auditable toolkit that pairs a reference implementation of the (additive)
fingerprint-copy attack with the \emph{triangle test} detector and a
cohort-calibrated decision procedure, so that practitioners can quantify how
fragile a PRNU attribution is against this attack on their own pipelines. On a
controlled multi-camera Monte-Carlo benchmark the attack reaches an arbitrary
target PCE at a mean PSNR of $59.2$\,dB (visually lossless), yet the triangle
test separates forgeries from innocent images with an area under the ROC curve
of $0.999$, detecting every forgery at a cohort-calibrated $5\%$ false-positive
operating point. We discuss why a fixed significance cut-off is \emph{not} safe
in casework, the limitations of synthetic validation, and the open problem of
extending explainable detection to the modern, GAN-based generation of
fingerprint-transfer attacks. The toolkit is released under the MIT license
with an explicit dual-use statement.
\end{abstract}

\section{Introduction}
Digital images carry a weak but persistent multiplicative noise pattern induced
by manufacturing imperfections in the sensor, the Photo-Response Non-Uniformity
(PRNU)~\cite{lukas2006}. Because the pattern is unique to an individual sensor
and stable across a device's lifetime, it functions as a device fingerprint and
underpins source-camera identification in operational forensics~\cite{chen2008}.
The standard examination estimates a reference fingerprint $\K$ from a set of a
camera's images and tests a questioned image for the presence of $\K$ using a
shift-robust detection statistic, the Peak-to-Correlation Energy
(PCE)~\cite{goljan2008}. A PCE above a calibrated threshold is reported as
support for the proposition that the questioned image was acquired with that
camera.

The evidential weight of this conclusion rests on an implicit assumption: that
the only way for a camera's fingerprint to appear in an image is for the image
to have been taken with that camera. Goljan, Fridrich and Chen showed that this
assumption is false~\cite{goljan2011}. An adversary with access to the target
camera's images---for instance a person's publicly posted photographs---can
estimate $\K$ and \emph{paste} a scaled copy of it into an unrelated image. The
resulting forgery exhibits a high PCE against the target fingerprint and would,
to a PCE-only examiner, falsely attribute the image to the target device. In a
forensic setting the consequence is the fabrication of source-camera evidence
against a person.

The same work introduced a defence, the \emph{triangle test}, which exploits a
structural weakness of the attack: the copied fingerprint is statistically tied
to the specific images from which it was estimated, and that dependency can be
detected. Fifteen years on, the attack has migrated to deep generative
models~\cite{cozzolino2021spoc}, in step with the broader rise of
synthetic-media forensics~\cite{verdoliva2020}, while reference
implementations of the detector---and of the attack itself, for the purpose of
\emph{measuring} robustness---remain scattered and largely unavailable to
practitioners.

\paragraph{Contributions.}
\begin{enumerate}
  \item We release \code{prnucf}, an open, dependency-light Python toolkit that
  implements the full PRNU attribution pipeline (fingerprint estimation, PCE),
  the additive fingerprint-copy attack and its removal counterpart, and the
  triangle-test detector with a cohort-calibrated decision rule.
  \item We specify a practical, examiner-oriented detection procedure in which
  the triangle-test statistic of a questioned image is calibrated against a
  reference cohort of known-unrelated images, so that the false-positive rate is
  controlled by a population rather than by an assumed null.
  \item We provide a reproducible multi-camera Monte-Carlo benchmark quantifying
  the attack's success against a PCE examiner and the detector's separation
  power, and we report honestly on where a naive fixed threshold fails.
  \item We situate the classical additive attack within the current,
  deep-learning arms race and identify explainable detection of learned
  fingerprint transfer as the open problem.
\end{enumerate}
The toolkit is intended for robustness evaluation on one's own or consented
data. It deliberately omits target-acquisition tooling and any refinement whose
only purpose is to evade the detector (Section~\ref{sec:ethics}).

\section{Related Work}
\paragraph{PRNU attribution.}
Luk\'a\v{s} et al.~\cite{lukas2006} established sensor pattern noise as a
camera fingerprint; Chen et al.~\cite{chen2008} formalised maximum-likelihood
fingerprint estimation and the detection of origin and integrity; Goljan
et al.~\cite{goljan2008} introduced the PCE statistic and its false-alarm
calibration. These remain the backbone of operational practice, and recent work
continues to strengthen attribution with convolutional
networks~\cite{robustcnn2024} and to survey the field~\cite{sensors2025survey}.

\paragraph{Counter-forensics: the fingerprint-copy attack.}
Goljan, Fridrich and Chen~\cite{goljan2011} introduced both the fingerprint-copy
attack and the triangle test. The attack is additive: a copy of the estimated
fingerprint, scaled by image content, is added to a carrier image. Because the
copy is built from a specific set of the victim's images, each of those images
is correlated with the forgery beyond what a genuine shared-camera origin would
explain---the signal the triangle test measures.

\paragraph{Learned fingerprint transfer and anonymisation.}
The attack has since moved to generative models. \emph{SpoC}~\cite{cozzolino2021spoc}
uses a GAN to graft a target camera model's traces into an image, fooling CNN
camera-model identifiers and GAN-image detectors, including unseen ones, without
explicit fingerprint arithmetic---weakening precisely the statistical dependency
the triangle test relies on. Subsequent work extends fingerprint transfer across
both physical and AI-generated images~\cite{prnutransfer2025}, the setting that
overlaps most directly with synthetic-media (deepfake) forensics. On the
suppression side, camera-trace erasing and conditional adversarial camera-model
anonymisation~\cite{andrews2020} go beyond the inverse-flat-field and denoising
removal we include. Deep learning has likewise been brought to the defence of
attribution itself---learning a camera(-model) fingerprint directly, as in
Noiseprint~\cite{noiseprint2020}, or boosting sensor-based attribution with a
CNN~\cite{spncnn2019}---although such learned detectors raise their own
explainability questions. A 2024 survey of image and video anti-forensics and
counter-anti-forensics~\cite{survey2024} notes that implanted fingerprints still
leave detectable traces, but that counter-anti-forensic detectors remain narrow
and technique-specific and that the poor explainability of learned detectors is
a live admissibility concern in court---motivating the explainable,
statistically grounded detector we package here.

\section{Background}
\label{sec:background}
Let $\I$ denote a grayscale (or luminance) image. A denoising filter $F$ yields
the noise residual
\begin{equation}
  \W = \I - F(\I),
\end{equation}
which contains the PRNU term together with other noise. We use the wavelet-based
Wiener denoiser of~\cite{lukas2006}. Given $n$ images $\I_1,\dots,\I_n$ from one
camera, the maximum-likelihood reference fingerprint is
\begin{equation}
  \hat{\K} = \frac{\sum_{i=1}^{n} \W_i \, \I_i}{\sum_{i=1}^{n} \I_i^2},
  \label{eq:mle}
\end{equation}
followed by zero-meaning and a spectral Wiener step to suppress non-unique
artefacts (demosaicing, JPEG blockiness)~\cite{chen2008}. Attribution of a
questioned image $\I_q$ to the camera tests the presence of $\K$ in $\W_q$ via
the PCE of the circular cross-correlation between $\W_q$ and $\K\,\I_q$: a sharp
peak at zero shift, large relative to the energy of the correlation plane,
indicates common source. A PCE above a calibrated threshold $\tau$
(operationally, values of a few tens upward) is reported as support for common
origin.

\section{Threat Model}
We assume an adversary who (i) can collect a set $\mathcal{J}=\{\I_1,\dots,\I_N\}$
of images taken by a target camera $A$ (e.g.\ a person's public photographs),
and (ii) wishes to make an unrelated carrier image---taken by a different camera
$B$, or synthesised---attribute to $A$. The adversary estimates $\hat{\K}_A$ from
$\mathcal{J}$ via Eq.~\eqref{eq:mle} and injects it into the carrier. The
adversary's goal is a forgery whose PCE against $\hat{\K}_A$ exceeds the
examiner's threshold $\tau$ while remaining visually unaltered. The examiner, in
turn, possesses or can assemble $\mathcal{J}$ (or a superset of likely harvested
images) and a reference cohort of images from unrelated cameras, and wishes to
decide whether $\I_q$ carries a fingerprint \emph{copied from} $\mathcal{J}$
rather than one acquired by genuine capture.

\section{Method}
\subsection{The additive fingerprint-copy attack}
Following~\cite{goljan2011}, the forgery is
\begin{equation}
  \I' = \I + \alpha \, \I \odot \K,
  \label{eq:inject}
\end{equation}
where $\odot$ is the element-wise product and $\alpha>0$ controls injection
strength. The multiplicative coupling to $\I$ mirrors the physical PRNU model.
$\alpha$ trades off detectability-by-PCE against visible distortion. Our
implementation optionally solves for the $\alpha$ that attains a target PCE by
bounded bisection, so an evaluator can ask ``what strength is needed to cross my
threshold, and is the result still clean?''. We deliberately implement only this
published additive model and add no refinement aimed at defeating the detector.
A removal counterpart, $\I/(1+\alpha\K)$ (inverse flat-fielding) or residual
denoising, is included for the attribution-denial / anonymisation scenario.

\subsection{Triangle-test detection}
Let $c(\cdot,\cdot)$ denote the normalised correlation of two residuals. For the
questioned image $Y$ and the candidate source pool $\mathcal{J}$, form, over all
triples $\{i<j\}$,
\begin{equation}
  T = \operatorname*{mean}_{i<j}\; c(W_Y,W_i)\,c(W_Y,W_j)\,c(W_i,W_j).
  \label{eq:triangle}
\end{equation}
If $Y$ was not built from $\mathcal{J}$, its correlations $c(W_Y,W_i)$ carry only
whatever fingerprint $Y$ and $\I_i$ genuinely share and are uncorrelated with the
pool's internal structure $c(W_i,W_j)$; $T$ then sits at a baseline level. If $Y$
was forged by copying a fingerprint estimated from $\mathcal{J}$, each
$c(W_Y,W_i)$ gains a component proportional to $\I_i$'s contribution to
$\hat{\K}$, itself tied to $c(W_i,W_j)$, pushing $T$ above baseline.

\subsection{Cohort calibration}
Rather than assume an analytic null for $T$, we calibrate against a reference
cohort $\mathcal{R}$ of images of known-unrelated origin, each scored by
Eq.~\eqref{eq:triangle} against $\mathcal{J}$. Using the cohort's robust location
and scale (median and $1.4826\times$MAD), we report a $z$-score
\begin{equation}
  z = \frac{T_Y - \operatorname{med}(\{T_r\}_{r\in\mathcal{R}})}
           {1.4826\,\operatorname{MAD}(\{T_r\}_{r\in\mathcal{R}})},
\end{equation}
and flag $Y$ as a suspected copy-forgery when $z$ exceeds an operating threshold.
Crucially, the threshold is an \emph{operating point} chosen from the cohort, not
a fixed significance value: as Section~\ref{sec:exp} shows, the innocent
distribution of $z$ has a heavy tail, so a naive cut-off such as $z=3$ is unsafe.

\section{Experiments}
\label{sec:exp}
\paragraph{Protocol.}
We validate on synthetic sensors so the ground truth of ``who took the image'' is
known exactly. Each of $N=100$ trials draws two i.i.d.\ multiplicative sensor
fingerprints (cameras $A$, $B$; $128\times128$; PRNU std $0.03$) and smooth
scene content with additive read noise. The attacker estimates $\hat{\K}_A$ from
a pool of $10$ images of $A$; a reference cohort of $20$ images from distinct
unrelated cameras is drawn. A carrier from camera $B$ is forged by
Eq.~\eqref{eq:inject}, with $\alpha$ tuned so the forgery reaches a fixed PCE
decision threshold of $60$. We then score, with the triangle test, both the
forgery and an innocent (un-forged) image from camera $B$, each against the
$A$-pool and calibrated on the cohort. Code and seed are released for exact
reproduction.

\begin{figure}[t]
\centering
\includegraphics[width=0.72\textwidth]{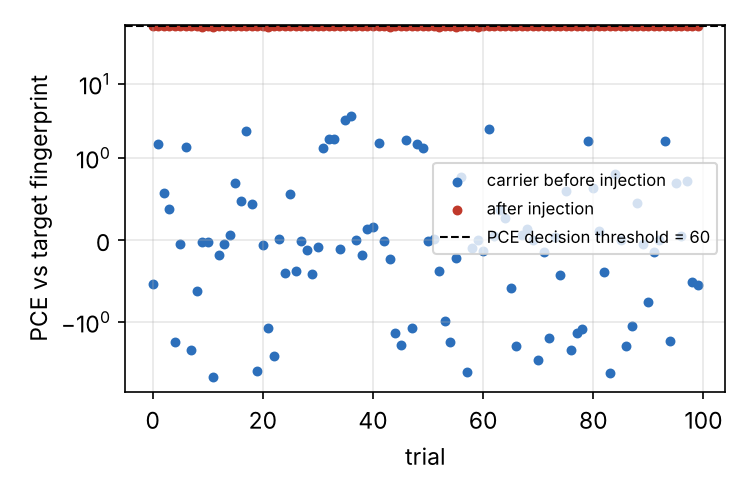}
\caption{PCE of each trial's carrier against the target fingerprint, before
(blue) and after (red) injection. The attack drives every carrier---initially
uncorrelated with the target camera---to the examiner's decision threshold
(dashed), at $59$\,dB PSNR.}
\label{fig:pce}
\end{figure}

\paragraph{Attack success.}
The carrier's PCE against $\hat{\K}_A$ is $\approx 0$ before injection and is
driven to the decision threshold of $60$ after injection, at a mean injection
strength $\bar\alpha = 1.02\pm0.24$ and a mean PSNR of $59.2\pm2.9$\,dB between
carrier and forgery---i.e.\ no visible alteration (Fig.~\ref{fig:pce}). A
PCE-only examiner would therefore attribute every forgery to camera $A$.

\begin{figure}[t]
\centering
\begin{subfigure}{0.48\textwidth}
  \includegraphics[width=\linewidth]{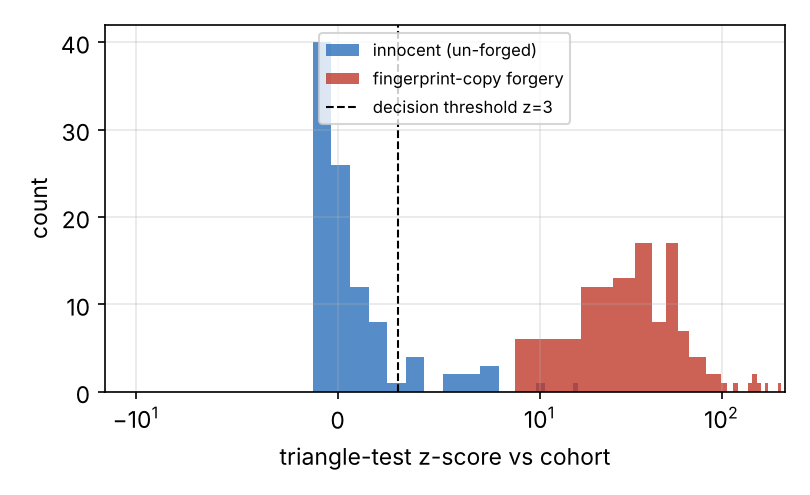}
  \caption{Triangle-test $z$-scores.}
  \label{fig:z}
\end{subfigure}\hfill
\begin{subfigure}{0.48\textwidth}
  \includegraphics[width=\linewidth]{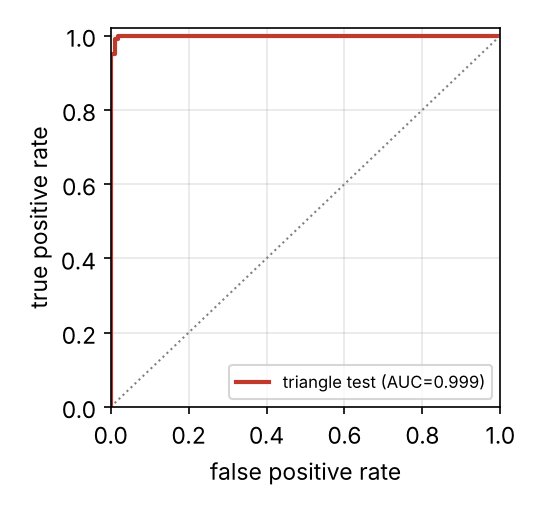}
  \caption{ROC of the detector.}
  \label{fig:roc}
\end{subfigure}
\caption{Detection performance over $100$ trials. (a) Innocent images cluster
near $z=0$ with a heavy tail; forgeries are pushed to $z\in[9,200]$. The dashed
line marks a naive $z=3$ cut-off, which is too aggressive. (b) The triangle test
separates forged from innocent with AUC $0.999$.}
\end{figure}

\paragraph{Detection.}
The triangle-test $z$-scores separate the two populations almost completely
(Fig.~\ref{fig:z}): innocent images have $\bar z = 0.9\pm2.7$ while forgeries
have $\bar z = 51.7\pm35.9$ with a minimum of $8.8$. The detector attains an area
under the ROC curve of $0.999$ (Fig.~\ref{fig:roc}). Table~\ref{tab:results}
summarises operating points: at a cohort-calibrated $5\%$ false-positive rate the
triangle test detects \emph{all} forgeries ($100\%$ TPR, threshold $z\approx6.9$),
and $99\%$ at a $1\%$ false-positive rate ($z\approx10.2$).

\paragraph{Why a fixed cut-off is unsafe.}
The innocent $z$-distribution is not standard normal: it is heavy-tailed, with a
maximum of $16.1$ across $100$ trials. A naive $z=3$ cut-off, though it catches
every forgery, misclassifies $13\%$ of innocent images. The separation is real
(AUC $0.999$), but the \emph{operating threshold must be set from the cohort},
not assumed---a point of direct practical consequence for casework, where a
false ``forgery'' finding is as damaging as a missed one.

\begin{table}[t]
\centering
\caption{Monte-Carlo results ($N=100$ trials, $128\times128$, pool $=10$,
cohort $=20$, target PCE $=60$).}
\label{tab:results}
\begin{tabular}{lr}
\toprule
Quantity & Value \\
\midrule
Carrier PCE before injection (mean) & $\approx 0$ \\
Forgery PCE after injection (mean) & $60.0$ \\
Injection strength $\alpha$ (mean $\pm$ sd) & $1.02 \pm 0.24$ \\
Carrier--forgery PSNR (mean $\pm$ sd) & $59.2 \pm 2.9$\,dB \\
\midrule
Triangle-test $z$, innocent (mean $\pm$ sd) & $0.9 \pm 2.7$ \\
Triangle-test $z$, forgery (mean $\pm$ sd) & $51.7 \pm 35.9$ \\
Triangle-test $z$, forgery (min) & $8.8$ \\
ROC AUC & $0.999$ \\
TPR at $5\%$ FPR (threshold $z$) & $1.00$ \;($6.9$) \\
TPR at $1\%$ FPR (threshold $z$) & $0.99$ \;($10.2$) \\
FPR at naive $z=3$ cut-off & $0.13$ \\
\bottomrule
\end{tabular}
\end{table}

\paragraph{Limitations of synthetic validation.}
Synthetic sensors give exact ground truth but idealised statistics. Real
residuals carry demosaicing and compression artefacts, scene content is more
diverse, and fingerprint estimates are noisier; the heavy innocent tail we
observe is partly an artefact of correlated synthetic scene content and a finite
cohort, and would change on real data. Absolute thresholds reported here are
\emph{not} transferable to casework: they must be re-derived on device-matched
ground-truth imagery. The experiment establishes the \emph{mechanism}---PCE is
foolable, the triangle test recovers the forgery with high separation---not
operational cut-offs.

\section{Ethical Considerations and Responsible Release}
\label{sec:ethics}
This work publishes an attack. We do so because the attack has been in the
peer-reviewed literature since 2011~\cite{goljan2011}; withholding a clean
reference implementation does not deny the capability to a motivated adversary,
it only denies examiners a means to test whether their own conclusions are robust
to it. We mitigate the dual-use risk by construction: the toolkit ships the
attack \emph{together with} its detector; the injection implements only the
already-published additive model and contains no refinement tuned to evade the
triangle test or any deployed system; there is no target-acquisition tooling; and
operation is intended on one's own or consented data. The release carries an
explicit ethics statement to this effect. Using fingerprint injection to make a
questioned image attribute to a camera that did not take it---i.e.\ to fabricate
source-camera evidence against a person---is evidence tampering and, in most
jurisdictions, a crime. A triangle-test result is moreover one line of evidence:
a flag indicates a questioned image shares a fingerprint statistically tied to a
specific image pool; it is not, by itself, proof of intent or identity, and must
be weighed with provenance and the processing chain, reporting uncertainty to the
same standard as the PCE conclusion it protects.

\section{Conclusion}
A high PCE is strong but not conclusive evidence of camera origin: the
fingerprint-copy attack forges it at visually lossless quality. The triangle
test, calibrated against a reference cohort, recovers the forgery with an ROC AUC
of $0.999$ on our benchmark, provided the operating threshold is set from the
cohort rather than fixed. We release \code{prnucf} so that practitioners can run
this evaluation on their own pipelines. The open problem is the modern,
GAN-based generation of fingerprint-transfer attacks~\cite{cozzolino2021spoc,
prnutransfer2025}, which weaken the statistical dependency the triangle test
exploits: an equally explainable, court-admissible detector for learned
injection is, in our view, the most valuable next step---and a natural extension
of the toolkit presented here.

\section*{Reproducibility}
The toolkit, the experiment script, and the random seed used for all figures and
tables are released under the MIT license at the project repository. The complete
benchmark in Section~\ref{sec:exp} runs in minutes on a single CPU.

\section*{Acknowledgements}
The author used an AI assistant (Anthropic Claude) for part of the software
implementation and manuscript drafting; all methods, experiments, and claims were
specified, reviewed, and verified by the author, who takes full responsibility
for the content.

\end{document}